# Artefact-removal algorithms for Fourier domain Quantum Optical Coherence Tomography

Sylwia M. Kolenderska[1*†] and Maciej Szkulmowski[2**†]

[1]The Dodd-Walls Centre for Photonic and Quantum Technologies, Department of Physics, University of Auckland, Auckland 1010, New Zealand

[2]Institute of Physics, Faculty of Physics, Astronomy and Informatics, Nicolaus Copernicus University in Torun, Grudziądzka 5, 87-100 Toruń, Poland

[†]These authors contributed equally to this work
[*]skol745@aucklanduni.ac.nz
[**]maciej.szkulmowski@fizyka.umk.pl

**Abstract**

Quantum Optical Coherence Tomography (Q-OCT) is a non-classical equivalent of Optical Coherence Tomography and is able to provide a twofold axial resolution increase and immunity to resolution-degrading dispersion. The main drawback of Q-OCT are artefacts which are additional elements that clutter an A-scan and lead to a complete loss of structural information for multilayered objects. Whereas there are successful methods for artefact removal in Time-domain Q-OCT, no such scheme has been devised for Fourier-domain Q-OCT (FdQ-OCT), although the latter modality – through joint spectrum detection – outputs a lot of useful information on both the system and the imaged object. Here, we propose two algorithms which process a Fd-Q-OCT's joint spectrum into an artefact-free A-scan. We present the theoretical background of these algorithms and show their performance on computergenerated data. The limitations of both algorithms with regards to the experimental system and the imaged object are discussed.

## Introduction

Optical Coherence Tomography (OCT) has become an important tool in medicine [9] because it enables visualisation of internal structures of biomedical objects non-invasively and on a micrometre scale. It is based on an interferometric measurement of light's time of flight and is performed by axially translating a mirror in a reference arm as in Time-domain OCT [4] or by keeping the mirror fixed and measuring light's spectrum as in Fourier-domain OCT [2].

Quantum Optical Coherence Tomography (Q-OCT) [1] is a non-classical counterpart of OCT using the quantum nature of light. The core of Q-OCT is quantum interference of entangled photon pairs occurring in a Hong-Ou-Mandel interferometer. The photon pairs are created in a nonlinear crystal at the input of the interferometer: one of the photons penetrates the object in the object arm and the other photon is reflected from a mirror in the reference arm. They both overlap at a beamsplitter and photodiodes located at the beam-splitter's two output ports measure the coincidence of the photons' simultaneous arrival. As in the case of traditional OCT, Quantum OCT can be performed in two ways. In Time-domain Q-OCT (Td-Q-OCT) [8], a depth profile of the object – an A-scan – is obtained by axially translating the reference mirror and performing the coincidence rate measurement. In Fourier-domain QOCT (Fd-Q-OCT) [6], the mirror is fixed and the coincidence measurement is done together with wavelength discrimination producing a two-dimensional joint spectrum. An A-scan is obtained by Fourier transforming one of the diagonals of the joint spectrum and provides many benefits: an increased axial resolution and immunity to resolution-degrading even orders of chromatic dispersion

Unfortunately, the same quantum effects that are responsible for Q-OCT's extraordinary features give also rise to its one huge drawback – artefacts. Artefacts are additional peaks or dips in an A-scan which do not directly relate to the structure of the object and effectively lead to scrambling of the whole depth profile. They were the main reason why until recently Q-OCT, despite its great potential, has not been exhaustively studied in terms of its further imaging capabilities.

In the last year, the interest in Q-OCT has been revived and the first successful scheme allowing for a reduction of artefacts was finally presented. A potential strategy to remove the artefacts was already proposed in early studies on Q-OCT [1]. Having noticed that slight changes in the central frequency of the pump light used to generate entangled photon pairs make the artefacts in the resulting Td-Q-OCT depth profile transition from a peak to a dip and vice versa, the authors of Ref.[1] suggested that the artefacts can be entirely removed by averaging depth profiles taken for multiple pump frequencies. Seventeen years later in their 2019 paper, Graciano et al. [3] showed it experimentally by using a spectrally broadband light source as a pump. Because a broadband pump could be viewed as a sum of different central frequencies, the resultant depth profile is basically a coherent integration of depth profiles that would be created if each of these frequencies were used separately to produce a depth profile.

Very basic artefact removal numerical algorithms were also proposed for Fd-Q-OCT [6]. By being applied to Fourier transforms of the diagonals of two-dimensional joint spectra, they are far from being universal and could only be used for well-defined objects. Here, we present two algorithms which are applied directly to the joint spectrum and are able to reduce or completely suppress artefacts regardless of the type of an object. We compare the performance of these algorithms on Fd-Q-OCT signals numerically synthesised for different kinds of objects.

**Signal in Fd-Q-OCT**

The signal acquired in Fourier-domain Q-OCT (Fd-Q-OCT) [6] is a two-dimensional joint spectrum and is expressed in the following mathematical form:

$$C_{Fd-Q-OCT} = |\phi(\omega_\alpha, \omega_\beta)|^2 \left(|f(\omega_\alpha)|^2 + |f(\omega_\beta)|^2 - 2Re\{f(\omega_\alpha)f^*(\omega_\beta)\}\right) \qquad (1)$$

where $\omega_\alpha$ and $\omega_\beta$ are the frequencies of photons in a pair which add up to the frequency $2\omega_0$ of the pumping laser, $|\phi(\omega_\alpha, \omega_\beta)|^2$ is a two-dimensional joint spectral profile of the photon pairs, and $f$ is an object's transfer function which describes the phase delays which the object imparts on the light. $f$ is responsible for the appearance of fringes in the signal.

In the Fd-Q-OCT signal, the transfer function takes two different forms, each contributing to different elements in the A-scan after Fourier transformation. The term $|f(\omega_\alpha)|^2 + |f(\omega_\beta)|^2$ will generate stationary artefact peaks located at a fixed distance from zero optical path difference (OPD, 0 OPD corresponds to the zero point of the abscissa axis of the A-scan). For each artefact of this type, its distance will be equal to the distance between a pair of interfaces or scattering centres this artefact is related to. The last term in expression (1), $f(\omega_\alpha)f^*(\omega_\beta)$, will lead to peaks representing object's dispersion-cancelled and resolution-doubled structure as well as another type of an artefact: an instationary one which appears midway between two interfaces.

An example of a joint spectrum for an object consisting of three interfaces is presented in Fig. 1a with the X and Y axes being the optical frequency of the photons, $\omega_\alpha$ and $\omega_\beta$. The joint spectrum can be transformed to be presented in terms of the central frequency, $\omega_0$, and the frequency detuning from the central frequency, $\omega'$, where

$$\omega_\alpha = \omega_0 + \omega', \qquad (2)$$
$$\omega_\beta = \omega_0 - \omega', \qquad (3)$$

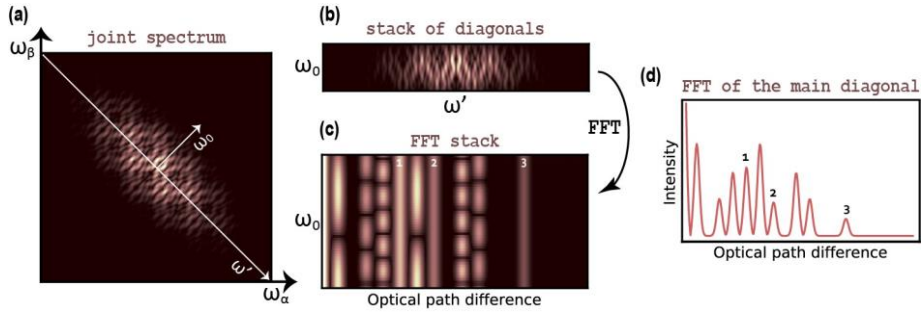

Figure 1: (a) The joint spectrum measured by Fd-Q-OCT is a function of the frequencies of photons, $\omega_\alpha$ and $\omega_\beta$. (b) The diagonals of the joint spectrum (most central 50 of them) are extracted along the direction of axis $\omega_0$ and put one on top of another to form a stack of diagonals. Each such diagonal spectrum represents a different central frequency $\omega_0$ (so a different frequency in the pump laser). (c) The rows of the stack of diagonals – when Fourier transformed – create an FFT stack which visualises differences between a solid-line structural peaks and intensity-varying artefact peaks. (d) A single row of an FFT stack – here, the FFT stack's central row which corresponds to a Fourier transform of the main diagonal of the joint spectrum – provides an A-scan cluttered with artefacts. 1, 2 and 3 number the structural peaks representing the object.

Effectively, this transformation leads to the extraction of diagonals of the joint spectrum and putting them one on top of another. Such a stack of diagonals is depicted in Fig. 1b (50 central diagonals in this case). Through the relationships (2) and (3), each such diagonal represents a dispersion-cancelled and resolution-doubled structure of the object. Since every diagonal corresponds to a different central frequency $\omega_0$ and consequently, to a different pump laser frequency $2\omega_0$, a joint spectrum broad in the anti-diagonal direction is obtained with spectrally broadband pump lasers. Broad anti-diagonal width of a joint spectrum is critical in Fd-Q-OCT – it enables extraction of a wider range of diagonals which – when Fourier transformed – provide a set of A-scans with enough information to distinguish artefact and structural peaks. Fig. 1c presents such a set of A-scans, an FFT stack. It can be seen in there that the height of the structural peaks remains constant with $\omega_0$ while the height of the artefact peaks oscillates. As mentioned before, there is a pair of artefacts for every two interfaces in the object's structure and each such pair oscillates with a different frequency which depends on the thickness and optical parameters of the layer (for more details see either [6] or [7]).

One of the A-scans from the FFT stack is presented in Fig. 1d and shows that a single diagonal is not enough to retrieve the true depth structure, because it is completely scrambled by artefacts.

Here and in the rest of this paper, an Fd-Q-OCT system was simulated which uses entangled photon pairs centred around 1560 nm and with total spectral bandwidth of 180 nm. The joint spectrum was assumed to be 256 by 256 points and enables the imaging range of 0.86 mm. The axial resolution of the simulated system is 5.6 $\mu$m.

## Results

### Complex averaging of the diagonals

In the first algorithm, the diagonals of the joint spectrum are extracted, recalculated to their complex representations using Hilbert transformation, then added up with proper weights and divided by their total number. This is an analogous procedure to the one proposed by Jensen et al. [5] for artefact removal in Intensity Correlation Spectral Domain OCT (ICA-SDOCT). ICA-SD-OCT is a quantum-mimic OCT method able to reproduce the advantageous features of Q-OCT – dispersion cancellation and resolution enhancement – using standard spectral-domain OCT data. As it was shown elsewhere [7], ICA-SD-OCT method processes an OCT spectrum into what can be viewed as the diagonals of Fd-Q-OCT's joint spectra. It recreates all the elements of joint spectrum's diagonals: dispersion-cancelled and resolution-doubled structure as well as stationary and instationary artefact peaks, but also adds new oscillatory artefact terms. Due to these additional artefacts, the correspondence of these two techniques is not ideal. However, since both techniques are based on analogous mechanisms (spectral correlation of photons in Fd-Q-OCT, intensity correlation in quantum-mimic OCT), the solutions of quantum-mimic OCT are easily adapted for use in quantum OCT (and vice versa) as is the case here.

The averaging algorithm can also be viewed as a software equivalent of the hardware approach of Graciano et al. [3]. In their approach, a broadband pump laser is used to generate entangled photon pairs with a joint spectrum broad in the anti-diagonal direction. The time-domain detection effectively complex "averages" the diagonals of the joint spectrum and outputs the Fourier transform of the outcome (for detailed analysis of a Q-OCT signal see Ref.[1], [7]).

To visualise the effects of applying the complex averaging algorithm to Fd-Q-OCT data, we used the joint spectrum from Fig. 1a and extracted 50 most central diagonals from it which covered the central wavelength range ($\lambda_0$ in Fig. 2a, where $\lambda_0 = \frac{2\pi c}{\omega_0}$) of 35 nm. Fourier transformation of the main diagonal of the joint spectrum – the central row of the stack of diagonals (Fig. 1b) – gives an A-scan full of artefacts (Fig. 1d). When all the rows from the stack of diagonals are averaged using a Kaiser window with $\beta$ = 6 as a weighting function, we obtain a spectrum (Fig. 1c) that is Fourier transformed to an artefact-free A-scan (Fig. 1d). More examples and a comparison with the second algorithm is presented in the Section Comparison.

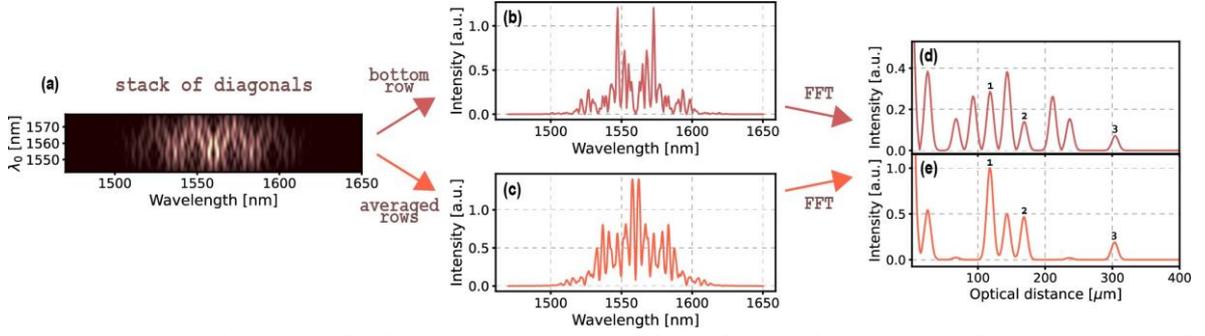

Figure 2: Using the example from Fig. 1, (a) a stack of diagonals consists of 50 most central diagonals of the joint spectrum and covers the central wavelength range of 35 nm. (b) The bottom row spectrum the stack – when Fourier transformed (FFT) – gives (d) an A-scan which is cluttered with artefacts. (c) Complex averaging of the spectra in the stack of diagonals outputs a spectrum which - when Fourier transformed - gives (d) an A-scan where the artefacts are reduced or completely suppressed. 1, 2 and 3 number the peaks representing the structure of the object.

The artefact reduction with this approach is the more successful, the more diagonals are used in it. Just as in the case of the equivalent algorithm of ICA-SD-OCT, a bigger number of diagonals corresponds to a bigger span of $\omega_0$'s (or $\lambda_0$'s shown in Fig. 2a). Since the artefact peaks oscillate as a function of $\omega_0$, a bigger span allows for more oscillations of artefact peaks (see Fig. 3c) and therefore, for a better artefact reduction as a result of the averaging of the diagonals. It was proved empirically in the article on ICA-SD-OCT that a complete artefact suppression is achieved when the averaging is done over the diagonals incorporating at least 5 full oscillations. This requirement poses a very strict limitation on how thin a layer can be for a given experimental setup so that the artefact suppression is absolute, and can be written as

$$\Delta z \geq 2S \frac{\lambda_{0c}^2}{\Delta \lambda_0} \frac{1}{n(\lambda_{0c})} \quad \text{or alternatively,} \quad \Delta z \geq 2S \frac{1}{2\pi c \Delta \omega_0} \frac{1}{n(\omega_{0c})}, \qquad (4)$$

where $S$ is the number of oscillations of an artefact, $\Delta z$ is the minimum thickness, $\Delta \lambda_0$ and $\Delta \omega_0$ are the total $\lambda_0$ and $\omega_0$ ranges covered by the diagonals, $\lambda_{0c}$ is the central $\lambda_0$, $\omega_{0c}$ is the central $\omega_0$, and $n(\lambda_{0c})$ and $n(\omega_{0c})$ are the refractive indices of the layer corresponding to $\lambda_{0c}$ and $\omega_{0c}$.

(4) also implies that the complex averaging algorithm will only be successful in removing the artefacts for layer's thicknesses which are on the order of the axial resolution if the antidiagonal spectral span of the joint spectrum ($\Delta \omega_0$) is 2S=2·5=10 times larger than the diagonal spectral span.

**Two-dimensional Fourier transform**

In the second approach, a two-dimensional Fourier transformation is applied to the two-dimensional joint spectrum given by (1). This operation transforms the signal from frequency space $(\omega_\alpha, \omega_\beta)$ to its representation in the inverse space $(z_\alpha, z_\beta)$ where all the components of the signal are better separated. To illustrate this, let us first consider a simple case of a layered object with two interfaces located at depths $z_1$ and $z_1 + z_2$. To account for dispersion of the layer, the wavenumber, $\beta = \beta(\omega)$, is expanded into Taylor series:

$$\beta(\omega) = \beta^{(0)} + \beta^{(1)}\omega + \beta^{(2)}\omega^2 + \beta^{(3)}\omega^3 + \cdots, \tag{5}$$

where $\beta^{(0)}$ is the wavenumber of light in air, $\beta^{(1)}$ – the inverse of the group velocity of light propagating in air, and $\beta^{(2)}$ and $\beta^{(3)}$ correspond to the second and third order dispersion exhibited by the layer.

After skipping the constant term $\beta_0$ that only influences the initial phase of the fringes, we obtain

$$\beta(\omega) = \beta^{(1)}\omega + \beta^{(2)}\omega^2 + \beta^{(3)}\omega^3 + \cdots = \beta^{(1)}\omega + \beta^D(\omega), \tag{6}$$

where $\beta^D$ incorporates all the dispersion coefficients equal to or higher than 2.

The transfer function, $f = f(\omega)$, describing the amplitude and phase changes introduced by the object can be written as:

$$f(\omega) = R_1 e^{iz_1 \beta_1(\omega)} + (1 - R_1) e^{iz_1 \beta_1(\omega)} R_2 e^{iz_2 \beta_2(\omega)} \tag{7}$$

where $R_1$ and $R_2$ are reflection coefficients of the interfaces of the object, and $\beta_1 = \beta_1(\omega)$ and $\beta_2 = \beta_2(\omega)$ are respectively the wavenumber of light propagating in the first medium, and the wavenumber of light propagating in the object. The above equation can be interpreted as follows. The light arrives at the first interface after having propagated in the first medium of thickness $z_1$ and part of it is reflected back, while the rest is transmitted forward and reaches the next interface after having propagated in the object of thickness $z_2$.

In the case of an object with $N$ reflective interfaces, (7) can be generalised into

$$f(\omega) = \sum_{n=1}^{N} R_n e^{iz_n \beta_n(\omega)} \prod_{m=1}^{n} (1 - R_{m-1}) e^{iz_{m-1} \beta_{m-1}(\omega)}, \tag{8}$$

where at zero-delay position $z_0 = 0$, there is no light attenuation and $R_0 = 0$.

In order to simplify the notation, let us define $S_n = R_n \prod_{m=1}^{n}(1 - R_{m-1})$. Then (8) can be rewritten as:

$$f(\omega) = \sum_{n=1}^{N} S_n e^{iz_n \beta_n(\omega)} \prod_{m=1}^{n} e^{iz_{m-1} \beta_{m-1}(\omega)} = \sum_{n=1}^{N} S_n e^{iz_n \beta_n(\omega) + i \sum_{m=1}^{n} z_{m-1} \beta_{m-1}(\omega)}. \tag{9}$$

Next, we expand the wavenumber and reorder the terms:

$$f(\omega) = \sum_{n=1}^{N} S_n e^{i\left(z_n \beta_n^{(1)} + i \sum_{m=1}^{n} z_{m-1} \beta_{m-1}^{(1)}\right)\omega + i\left(z_n \beta_n^D(\omega) + i \sum_{m=1}^{n} z_{m-1} \beta_{m-1}^D(\omega)\right)}. \tag{10}$$

After introducing optical distance $\tilde{z}_n = z_n \beta_n^{(1)} + i \sum_{m=1}^{n} z_{m-1} \beta_{m-1}^{(1)}$ and the dispersion-related phase distortion $\phi_n^D(\omega) = z_n \beta_n^D(\omega) + i \sum_{m=1}^{n} z_{m-1} \beta_{m-1}^D(\omega)$, we obtain:

$$f(\omega) = \sum_{n=1}^{N} S_n e^{i\tilde{z}_n \omega} e^{i\phi_n^D(\omega)}. \tag{11}$$

It can now be easily seen that each term is composed of two phase terms, namely the one responsible for the position of the reflecting layer and the one responsible for the dispersive broadening.

Now, the last term in the expression (1), $2Re\{f(\omega_\alpha)f^*(\omega_\beta)\} = M(\omega_\alpha, \omega_\beta)$ responsible for dispersion cancellation, resolution doubling and instationary artefacts can be expressed as:

$$M(\omega_\alpha, \omega_\beta) = 2\text{Re}\left\{\left(\sum_{n=1}^{N} S_n \exp(i\tilde{z}_n \omega_\alpha)\exp(i\phi_n^D(\omega_\alpha))\right)\left(\sum_{u=1}^{N} S_u \exp(-i\tilde{z}_n \omega_\beta)\exp(-i\phi_u^D(\omega_\beta))\right)\right\} =$$
$$= 2\text{Re}\left\{\sum_{n=1}^{N}\sum_{u=1}^{N} S_n S_u \exp(i(\tilde{z}_n \omega_\alpha - \tilde{z}_u \omega_\beta))\exp(i(\phi_n^D(\omega_\alpha) - \phi_u^D(\omega_\beta)))\right\} =$$
$$= 2\text{Re}\left\{\sum_{n=1}^{N}\sum_{u=1}^{N} S_n S_u P_{n,u}(\omega_\alpha, \omega_\beta) D_{n,u}(\omega_\alpha, \omega_\beta)\right\}.$$
(12)

where $P_{n,u}$ will be responsible for a placement of peaks in the inverse space $(z_\alpha, z_\beta)$, $D_{n,u}$ is a term responsible for dispersion effects.

The retrieval of the structural information is done by applying Fourier transformation twice to (12), once along $\omega_\alpha$ axis and once along $\omega_\beta$ axis. The resultant two-dimensional Fourier transform, $m$, represents the signal in terms of optical time delays which are recalculated to optical path differences, or thicknesses, $(z_\alpha, z_\beta)$:

$$m(z_\alpha, z_\beta) = \mathcal{F}\{M(\omega_\alpha, \omega_\beta)\} = \sum_{n=1}^{N}\sum_{u=1}^{N} S_n S_u \delta(z_\alpha - \tilde{z}_n, z_\beta + \tilde{z}_u) \otimes \Phi_{n,u}^D(z_\alpha, z_\beta), \qquad (13)$$

where $\Phi^D$ is the Fourier transform of $D$.

It can be seen from the delta function that the two-dimensional Fourier transform consists of peaks located at the positions $(\tilde{z}_n, -\tilde{z}_u)$, where the structural peaks are located at the diagonal for which $n = u$ and artefact peaks will be placed off-diagonally where $n \neq u$. Therefore, in order to reconstruct the positions of the layers without the parasitic terms it is sufficient to take the diagonal. However, one has to have in mind that the closer two interfaces are located, the closer to the diagonal their parasitic terms are and their impact on the diagonal starts to be visible. The convolution of the delta function with the dispersive term $\Phi^D$ leads to the distorting of the off-diagonal artefact peaks due to all the non-zero higher-order dispersion terms, and distorting of the structural peaks due to any non-zero even-order dispersion terms. Also, the distance between the structural peaks is increased which results in the resolution increase. Let's assume an object consisting of a single $\Delta\tilde{z}$-thick layer. In traditional OCT A-scan, this object's interfaces are positioned at depth $\tilde{z}$ and $\tilde{z} + \Delta\tilde{z}$. In Fd-Q-OCT, the same two interfaces will create two peaks located at $(\tilde{z}, -\tilde{z})$ and $(\tilde{z} + \Delta\tilde{z}, -(\tilde{z} + \Delta\tilde{z}))$ whose distance is $\sqrt{2}\Delta\tilde{z}$. Since this geometrical distance does not account for optical relationships between photons, it needs to be multiplied by $\sqrt{2}$. Consequently, the distance between the two peaks is $2\Delta\tilde{z}$ which means a twice better resolving power of the method. A detailed explanation of why the multiplication by $\sqrt{2}$ needs to be performed can be found in Supplementary Document.

In summary, the second algorithm exploits the modulatory character of artefacts in the anti-diagonal direction ($\omega_0$ axis in Fig. 2c). Since structural peaks do not exhibit such behaviour, two-dimensional Fourier transform of a joint spectrum (depicted in Fig. 3a) comprises structural peaks on its diagonal and peaks corresponding to artefacts positioned at some distance from it. The shape of the peaks in the transform is not uniform and reflects the shape of the joint spectrum: the joint spectrum is broad in the diagonal direction and narrow in the anti-diagonal direction, which results in peaks being narrow in the diagonal direction and broad in the anti-diagonal direction. The diagonal of the 2D Fourier transform presented in Fig. 3b does not incorporate any artefact peaks.

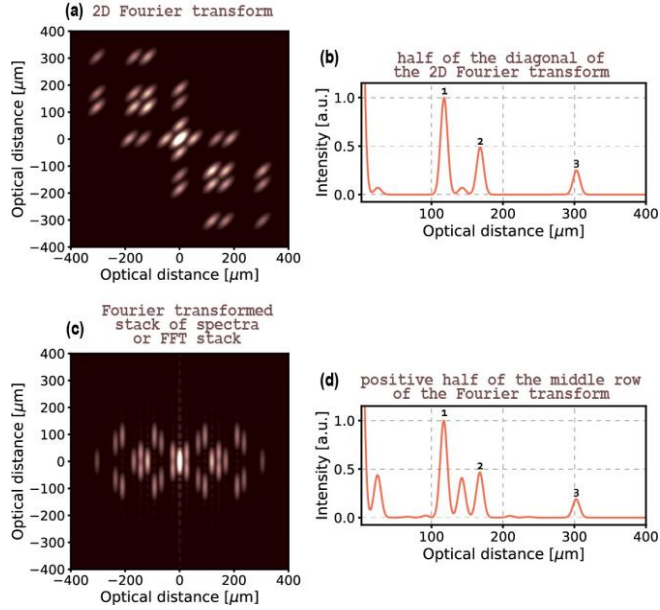

Figure 3: (a) 2D Fourier transformation is applied to the full joint spectrum from Fig. 1, then (b) its diagonal provides an A-scan where the artefacts are almost completely suppressed. (c) When the stack of spectra (from Fig. 1b or Fig. 2a) is 2D Fourier transformed or alternatively, when the FFT stack (from Fig. 1b) is Fourier transformed, (d) the middle row of the Fourier transform provides an A-scan where artefacts are reduced or completely suppressed. Note that the latter type of 2D Fourier transform approach gives a similar result as complex averaging of 50 central diagonals presented in Fig. 2. 1, 2 and 3 number the peaks representing the structure of the object.

Alternatively, a stack of diagonals or an FFT stack could be used to achieve a similar result. In the first case, a two-dimensional Fourier transform should be calculated and in the second case – a one-dimensional Fourier transform of each column. In both cases, the end result will be a two-dimensional Fourier transform which is a 45-degree rotation of the joint-spectrum-based transform (Fig. 3c). The middle row of such a transform is an artefact-free A-scan. Of course, in these two cases the width of a peak in the vertical (perpendicular to the rows) direction is broader than in the corresponding, anti-diagonal direction for a peak in the joint-spectrum-based transform, because the former use only a part of the joint spectrum. Because of that, some trails of the artefacts overlap the middle row area and consequently the artefact reduction is not absolute (Fig. 3d).

**Performance**

The performance of the two algorithms is summarised in Fig. 4. As it can be observed when Fig. 2 and Fig. 3 are compared, both approaches are equivalent when a sufficient number of diagonals is used in the complex averaging algorithm. Complex averaging of 50 diagonals corresponding to 35 nm $\lambda_0$ range and Fourier transformation of its output (a case presented in Fig. 2) gives the same result as performing 2D Fourier transformation on that very same stack and extracting the middle row (a case presented in Fig. 3c and d). In the example from Fig. 1, Fig. 2 and Fig. 3, to obtain equivalent results of applying 2D Fourier transformation to a full joint spectrum, 100 diagonals covering the $\lambda_0$ range of 70 nm are needed in the complex averaging algorithm as depicted in Fig. 4a-d. In general, so that the two algorithms are equivalent, the number of diagonals, $K$, in the complex averaging algorithm needs to cover the $\lambda_0$ span equal to at least the antidiagonal FWHM of a joint spectrum, $\Delta_{0.5}\lambda_0$. This requirement can be expressed as a ratio of the FWHM and the spectral resolution of the detection $\delta\lambda$:

$$K = \frac{\Delta_{0.5}\lambda_0}{\delta\lambda}. \tag{14}$$

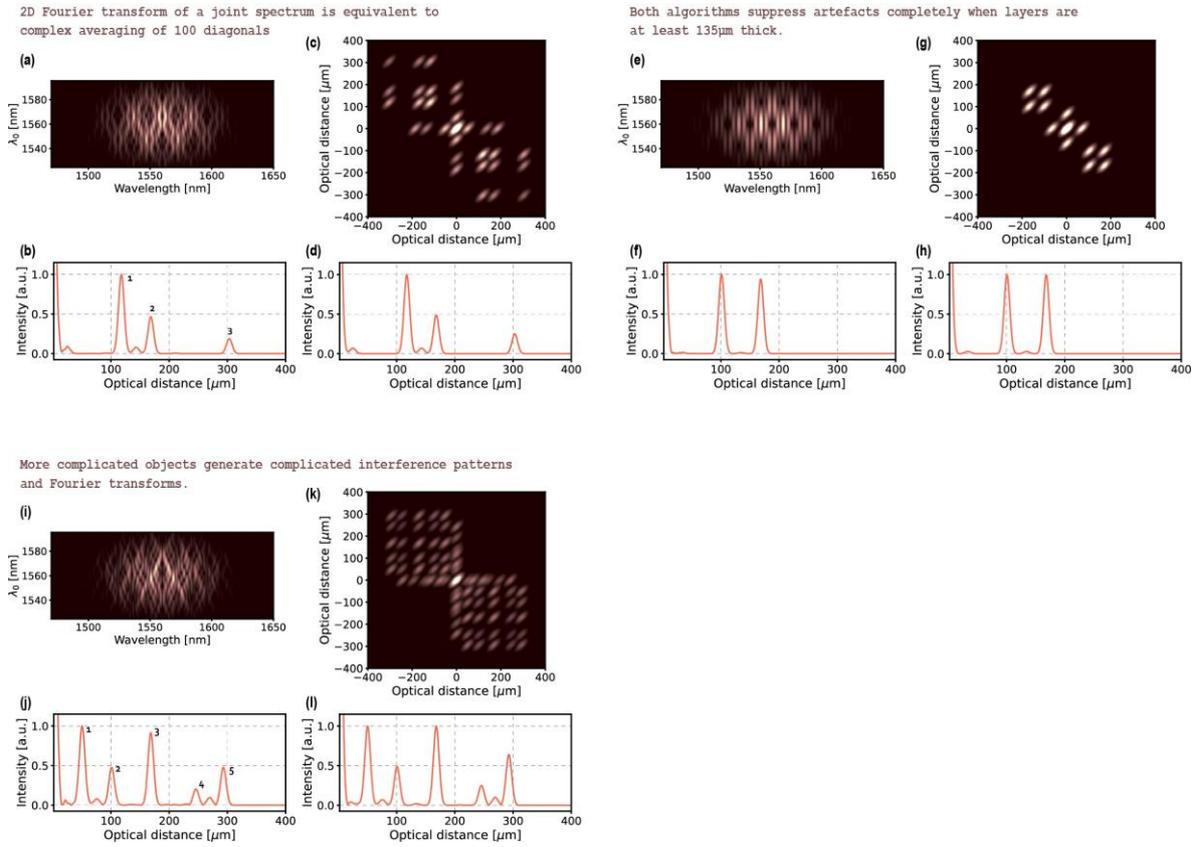

Figure 4: Complex averaging of (a) 100 diagonals extracted from a joint spectrum provides the same result (an A-scan with suppressed artefacts-(b)) as the 2D Fourier transform algorithm (the transform in (c) and the output A-scan in (d)). For both algorithms, the minimum thickness of a layer for which artefacts can be fully suppressed depends mainly on the optical parameters of the photons (see equation (4)) and in the experimental case simulated in this paper is 135 $\mu$m as it can be seen in the output A-scans of the complex averaging algorithm (f) and the output A-scan of the 2D Fourier transform algorithm (h) ((e) – the stack of diagonals, (g) – corresponding Fourier transform). More complicated objects generate more complicated spectral interference patterns on the stack of diagonals (i) and therefore, a more complicated Fourier transform (k), which when processed give A-scans with reduced artefacts ((j) – the output A-scans of the complex averaging algorithm, (f) – the output A-scan of the 2D Fourier transform algorithm).

As it was mentioned before, the algorithms successfully suppress artefacts only if a layer in the object is thick enough to generate a sufficient number of oscillations of the artefact peaks on an FFT stack. For the imaging scenario simulated in this paper, the minimum layer's thickness was determined to be 135 $\mu$m (Fig. 4e-h). The minimum layer's thickness can also be calculated using the relationship in (4).

A stack of diagonals and 2D Fourier transform for a signal corresponding to a more complicated object are presented in Fig. 4i and k together with the output of the two algorithms in Fig. 4j and l.

In the end, it should be noted that the A-scans output by the two algorithms are not perfectly dispersion-cancelled. This is due to the fact that they require several diagonal spectra from the joint spectrum, so although each such diagonal spectrum is a perfectly dispersion-cancelled signal for a given $\lambda_0$, their direct summation (as in the complex averaging algorithm) or integration (as in the Fourier transform algorithm) is not, as each diagonal will Fourier transform to an A-scan presenting slightly shifted optical distances. However, the dispersion broadening due to the 35 or 75 nm anti-diagonal bandwidths presented in this paper will be much smaller than the broadening that would occur as a result of a 180 nm bandwidth of a diagonal spectrum if that spectrum did not exhibit the dispersion cancellation feature.

## Summary

We have presented two algorithms for artefact reduction in Fourier-domain Quantum Optical Coherence Tomography (Fd-Q-OCT). The first one complex averages the diagonals extracted from the signal acquired in Fd-Q-OCT – the joint spectrum – and then applies Fourier transformation to those averaged diagonals. In the second algorithm, two-dimensional Fourier transformation is applied directly to the joint spectrum and the diagonal of the resultant Fourier transform is taken. The former approach is a reproduction of an algorithm for artefact removal in the classical counterpart of Fd-Q-OCT called ICA-SD-OCT [5]. What is more, both algorithms can be viewed as a software counterpart of the hardware-wise artefact removal in Td-Q-OCT [3]. 2D Fourier transform algorithm is a new approach for which we prove with theoretical calculations that indeed, the artefact-free A-scan is found on the diagonal of the transform. We show that both approaches are equivalent if a sufficient number of diagonals of the joint spectrum is used in the complex averaging algorithm. We provide mathematical expressions enabling calculation of this number and also, the calculation of the minimum thickness of a layer for which the artefacts will be fully suppressed.

## References

[1] Ayman F Abouraddy et al. "Quantum-optical coherence tomography with dispersion cancellation". In: *Physical Review A* 65.5 (2002), p. 053817.

[2] Adolph F Fercher et al. "Measurement of intraocular distances by backscattering spectral interferometry". In: *Optics communications* 117.1-2 (1995), pp. 43–48.

[3] Pablo Yepiz Graciano et al. "Interference effects in quantum-optical coherence tomography using spectrally engineered photon pairs". In: *Scientific Reports* 9.1 (2019), pp. 1–14.

[4] David Huang et al. "Optical coherence tomography". In: *science* 254.5035 (1991), pp. 1178–1181.

[5] Mikkel Jensen et al. "All-depth dispersion cancellation in spectral domain optical coherence tomography using numerical intensity correlations". In: *Scientific Reports* 8.1 (2018), p. 9170.

[6] Sylwia M Kolenderska, Frédérique Vanholsbeeck, and Piotr Kolenderski. "Fourier domain quantum optical coherence tomography". In: *Optics Express* 28.20 (2020), pp. 29576–29589.

[7] Sylwia M. Kolenderska and Piotr Kolenderski. *Intensity correlation OCT – a true classical equivalent of quantum OCT able to achieve up to 2-fold resolution improvement in standard OCT images*. 2021. arXiv: 2101.04826[physics.optics].

[8] Magued B Nasr et al. "Demonstration of dispersion-canceled quantum-optical coherence tomography". In: *Physical review letters* 91.8 (2003), p. 083601.

[9] Adam M Zysk et al. "Optical coherence tomography: a review of clinical development from bench to bedside". In: *Journal of Biomedical Optics* 12.5 (2007), p. 051403.

*Supplementary document*

# Two-dimensional Fourier transformation of a rotated joint spectrum in Fd-Q-OCT

To see why the additional multiplication by $\sqrt{2}$ is necessary, we consider a two-dimensional Fourier transformation applied along the axes parallel, $\omega_\parallel$, and perpendicular, $\omega_\perp$, to the main diagonal. This operation is analogous to a rotation of the coordinate system by π/4. In such a case, the new coordinates are given by:

$$\begin{bmatrix} \omega_\parallel \\ \omega_\perp \end{bmatrix} = \frac{\sqrt{2}}{2} \begin{bmatrix} 1 & -1 \\ 1 & 1 \end{bmatrix} \begin{bmatrix} \omega_\alpha \\ \omega_\beta \end{bmatrix}, \tag{S1}$$

where

$$\begin{aligned} \omega_\parallel &= \frac{\sqrt{2}}{2}(\omega_\alpha - \omega_\beta) & \omega_\alpha &= \frac{\sqrt{2}}{2}(\omega_\perp + \omega_\parallel) \\ \omega_\perp &= \frac{\sqrt{2}}{2}(\omega_\alpha + \omega_\beta) \quad \text{and} & \omega_\beta &= \frac{\sqrt{2}}{2}(\omega_\perp - \omega_\parallel) \end{aligned} \tag{S2}$$

The term $P_{n,u}$ responsible for positioning the peaks in a two-dimensional Fourier transform can be rewritten:

$$\begin{aligned} P_{n,u}(\omega_\alpha, \omega_\beta) &= \exp\left(i(\tilde{z}_n \omega_\alpha - \tilde{z}_u \omega_\beta)\right) = \\ &= \exp\left(i\frac{\sqrt{2}}{2}\left(\tilde{z}_n(\omega_\perp + \omega_\parallel) - \tilde{z}_u(\omega_\perp - \omega_\parallel)\right)\right) = \\ &= \exp\left(i\sqrt{2}\left(\frac{\tilde{z}_n + \tilde{z}_u}{2}\omega_\parallel + \frac{\tilde{z}_n - \tilde{z}_u}{2}\omega_\perp\right)\right) = \\ &= \exp(i\sqrt{2}\bar{z}_{nu}\omega_\parallel)\exp\left(i\frac{\sqrt{2}}{2}\Delta\tilde{z}_{nu}\omega_\perp\right) = \\ &= P_{n,u}(\omega_\parallel, \omega_\perp), \end{aligned} \tag{S3}$$

$\omega_\parallel$ and $\omega_\perp$, are related to geometrical coordinates and need to be expressed in terms of the central frequency $\omega_0$ and frequency detuning, $\omega'$, to truly represent the diagonals. Using the two first relationships of (S2) and the fact that $\omega_\alpha$ and $\omega_\beta$ represent negatively correlated photons, for which $\omega_\alpha = \omega_0 - \omega'$ and $\omega_\beta = \omega_0 + \omega'$, we obtain:

$$\begin{aligned} \omega_\parallel &= -\sqrt{2}\omega' \\ \omega_\perp &= \sqrt{2}\omega_0 \end{aligned} \tag{S4}$$

Using (S4) in (S3) gives:

$$P_{n,u}(\omega_\parallel, \omega_\perp) = \exp(-i\sqrt{2}\bar{z}_{nu}\omega')\exp(i\Delta\tilde{z}_{nu}\omega_0) = P_{n,u}(\omega', \omega_0) \tag{S5}$$

After two-dimensional Fourier transformation along $\omega'$ and $\omega_0$ we obtain:

$$m(z_\|, z_\perp) = \mathscr{F}\{P_{n,u}(\omega', \omega_0)\} = \delta\left(z_\| + 2\bar{\tilde{z}}_{nu}, z_\perp - \Delta\tilde{z}_{nu}\right) \tag{S6}$$

At the diagonal, for which $n = u$, $m(z_\|, z_\perp)$ simplifies to:

$$m_{\text{diag}}(z_\|, z_\perp) = \delta(z_\| + 2\tilde{z}_n, z_\perp) \tag{S7}$$

It can be clearly seen from (S7) that indeed only the components along the diagonal remain. Also, these remaining components represent peaks which are separated by twice the distance what leads to two-fold resolution increase.